\documentclass{PoS}

\title{PDF Fits at HERA}

\ShortTitle{PDF Fits at HERA}

\author{\speaker{Amanda Cooper-Sarkar}\thanks{On behalf of H1 and ZEUS Collaborations}\\
        Oxford University\\
        E-mail: \email{a.cooper-sarkar@physics.ox.ac.uk}}

\abstract{The HERAPDF1.0 PDF set, which was an NLO QCD analysis based on 
H1 and ZEUS combined 
inclusive cross section data from HERA-I, has been updated to HERAPDF1.5 
by including 
preliminary inclusive cross section data from HERA-II running. Studies have also 
been made by adding various 
other HERA data sets: combined charm data, combined low energy run data and 
H1 and ZEUS jet data. These data give information on the treatment of charm and 
the charm quark mass and  on the value of $\alpha_s(M_Z)$. 
The PDF analysis has also been extended to NNLO. The PDFs give a good 
description of Tevatron and early LHC data.
 }

\FullConference{The 2011 Europhysics Conference on High Energy Physics-HEP 2011,\\
		July 21-27, 2011\\
		Grenoble, Rhône-Alpes France}

\begin{document}
\section{Introduction}

The HERAPDF1.0 NLO 
parton distribution functions (PDFs) were extracted using the 
combined inclusive cross section data from the H1 and ZEUS collaborations 
taken at the HERA collider during the HERA-I running period 
1992-1997~\cite{h1zeus:2009wt}. 
These data come from neutral and charged current interactions from both
$e^+ p$ and $e^- p$ scattering. The combination of the H1 and ZEUS data sets 
takes into account the full correlated systematic uncertainties of the 
individual experiments such that the total uncertainty of the combined measurement is typically smaller than 
$2\%$, for $3 < Q^2 < 500$~GeV$^2$, and reaches $1\%$, for $20 <  Q^2 < 
100$~GeV$^2$. Additional preliminary HERA data has been added to the HERAPDF1.0 
NLO QCD analysis: HERA combined charm data have been 
added~\cite{charmcomb}; combined data from low energy running have been 
added~\cite{lowEcomb}; further combined inclusive cross-section data from 
the HERA-II running period 2003-2007 have been added (HERAPDF1.5)~\cite{herapdf15};
H1 and ZEUS jet data have been added (HERAPDF1.6)~\cite{herapdf16}; 
finally a fit which comprises all these data sets has been made (HERAPDF1.7).
The NLO fits have also been extended to NNLO for both HERAPDF1.0 and 1.5~\cite{herapdf15nnlo} and 
the NLO and NNLO PDFs have been confronted with Tevatron and LHC data  

\section{Results}
 
\begin{figure}[tbp]
\vspace{-0.5cm} 
\begin{tabular}{cc}
\includegraphics[width=0.35\textwidth]{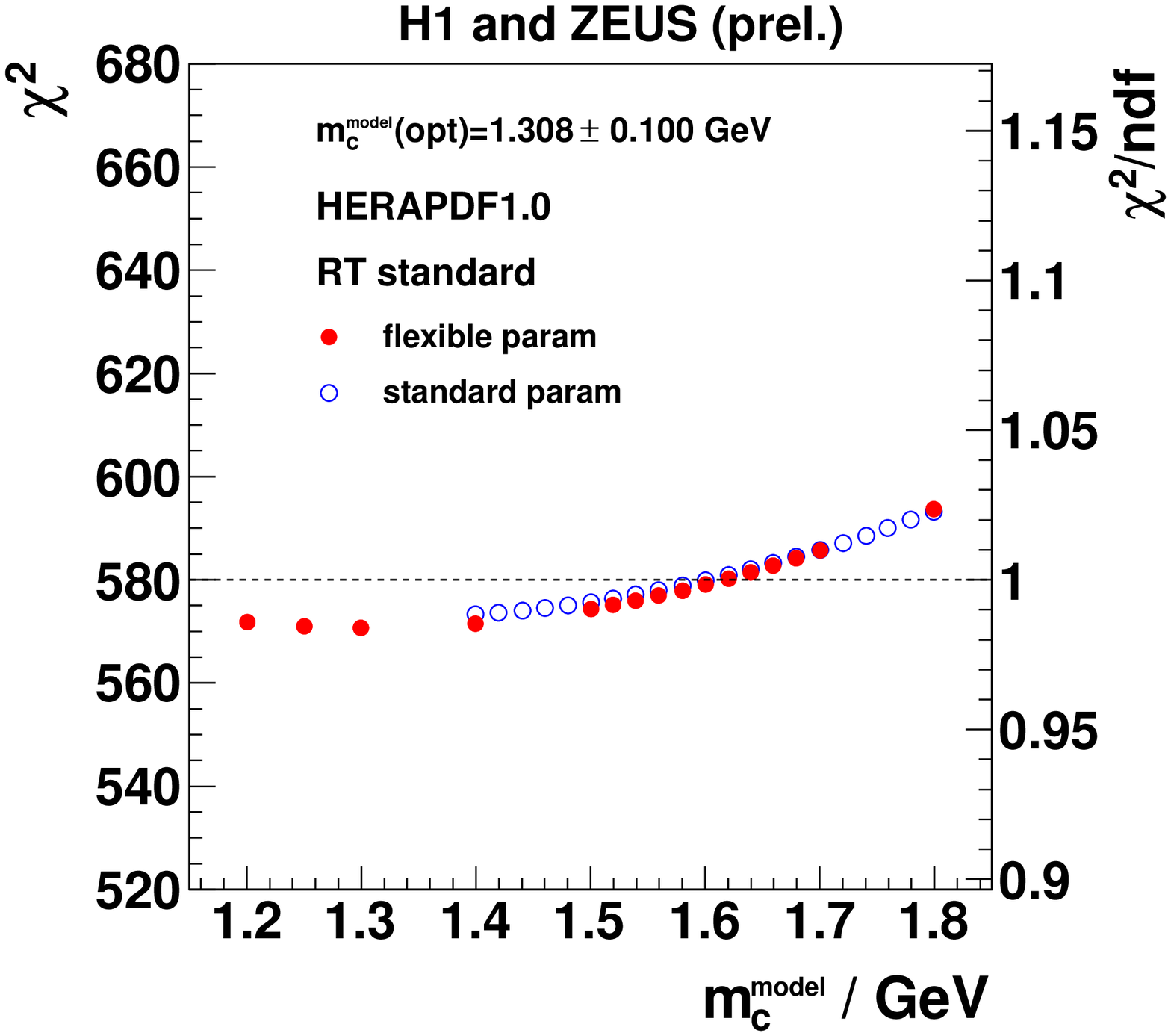} &
\includegraphics[width=0.35\textwidth]{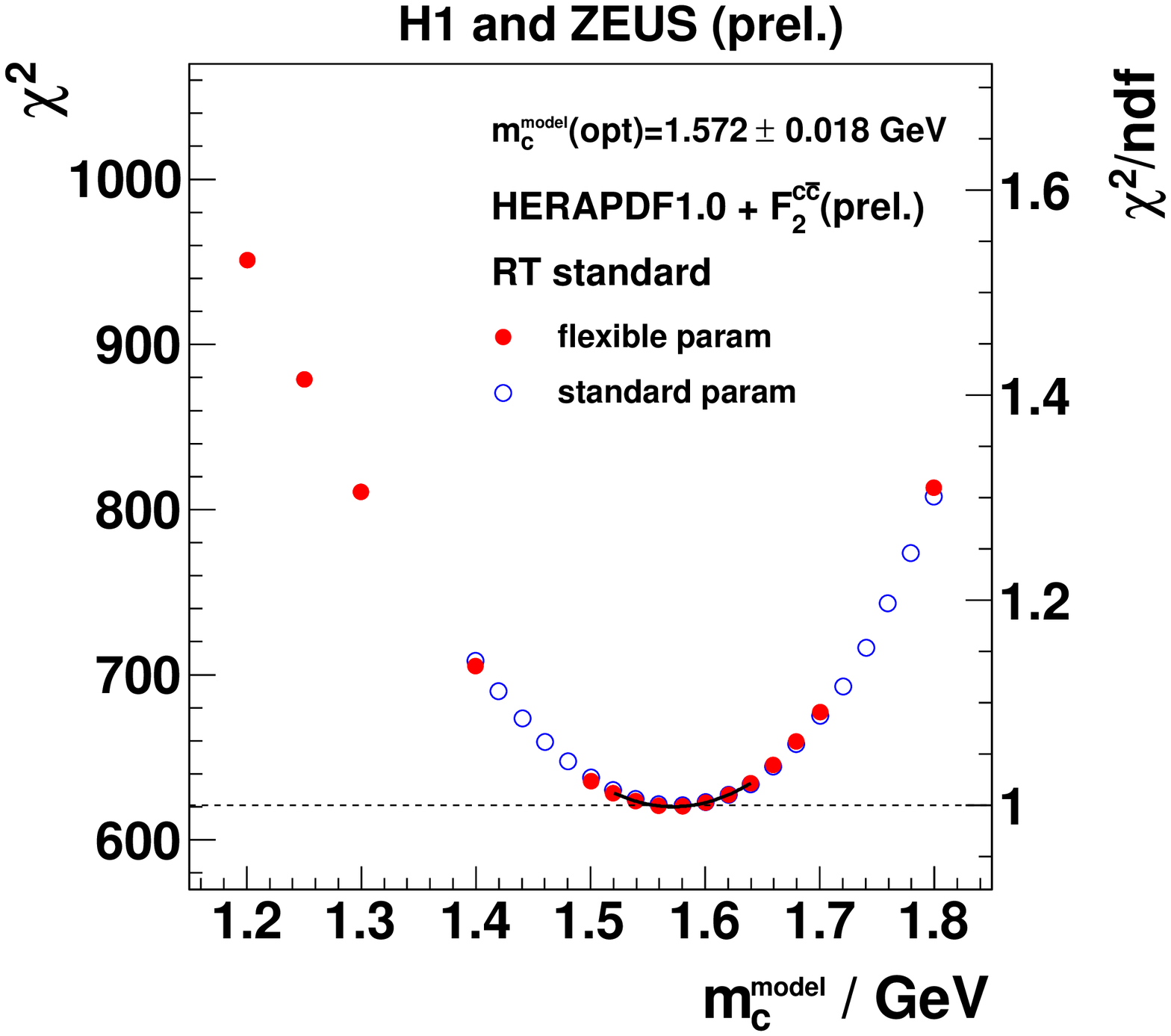}
\end{tabular}
\begin{tabular}{cc}
\includegraphics[width=0.35\textwidth]{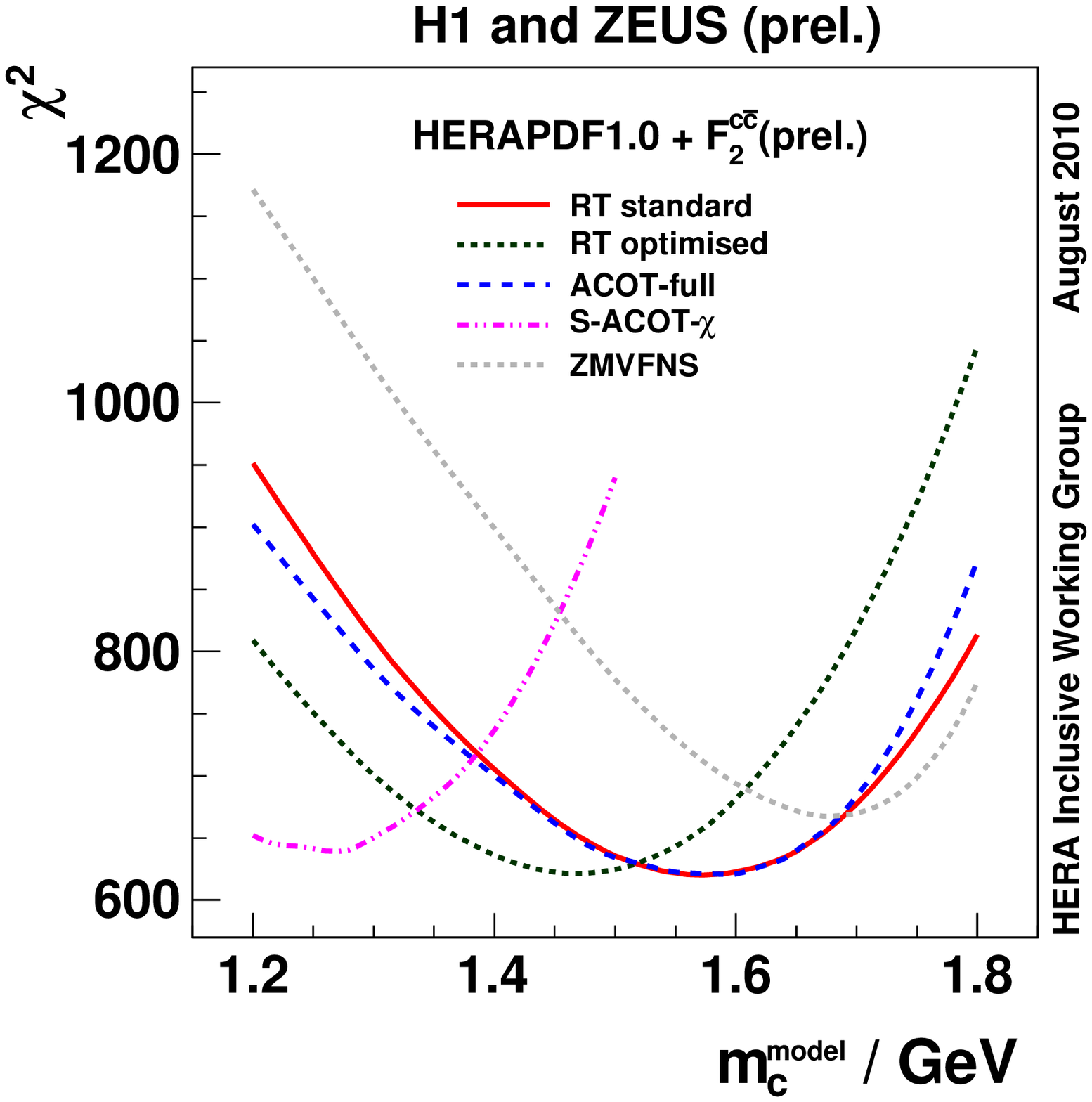} &
\includegraphics[width=0.35\textwidth]{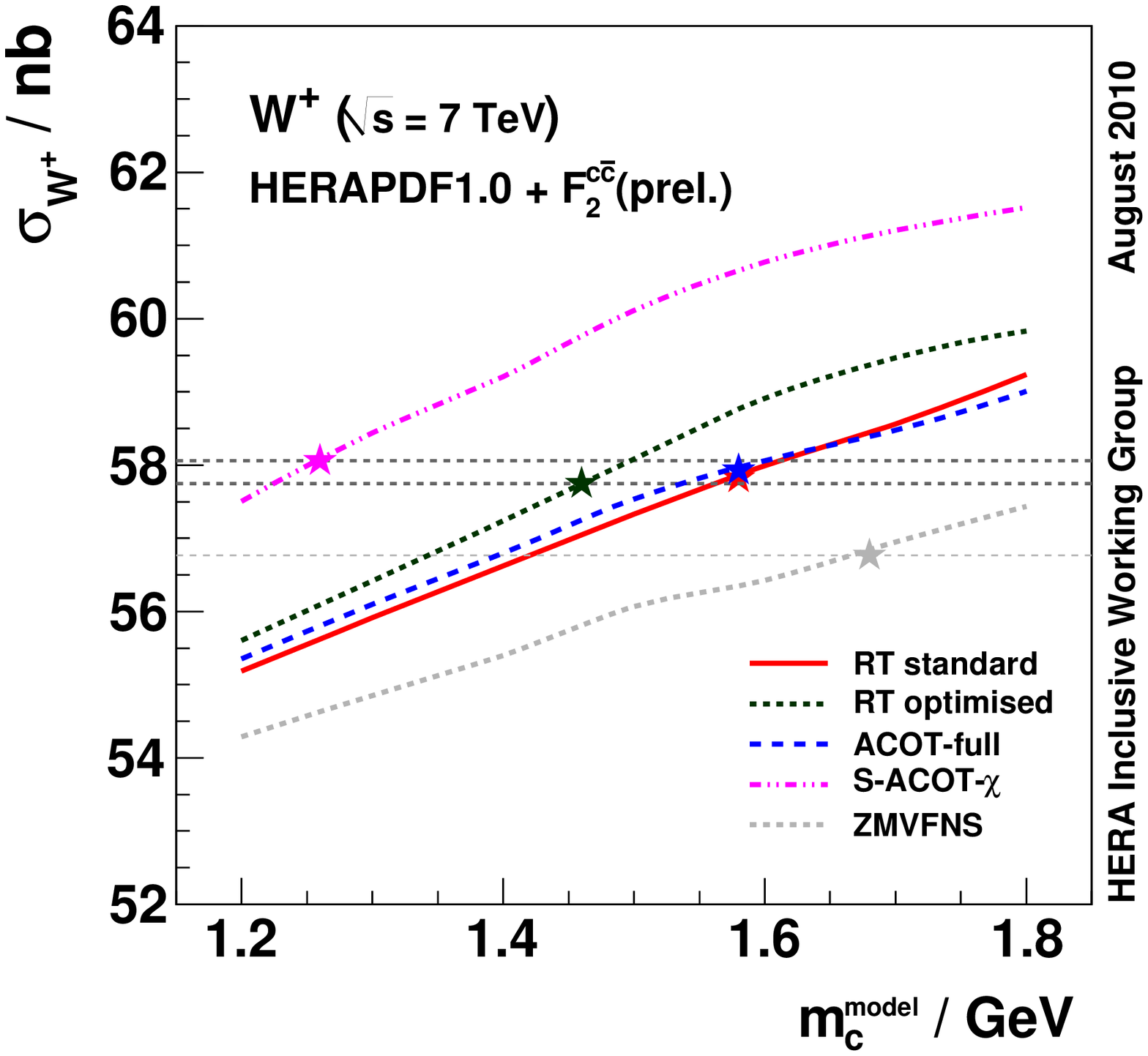}
\end{tabular}
\caption {The $\chi^2$ of the HERAPDF fit as a 
function of the charm mass parameter $m_c^{model}$. Top left; using the 
RT-standard heavy-quark-mass scheme, when only inclusive DIS data are included 
in the fit. Top right; 
using the RT-standard heavy-quark-mass scheme, 
when the data for $F_2^{c\bar{c}}$ are also included in the fit. Bottom left;
 using various heavy-quark-mass schemes, when the data for $F_2^{c\bar{c}}$
 are also included in the fit. Bottom right: predictions for the $W^+$ cross-sections at the LHC, as a 
function of the charm mass parameter $m_c^{model}$, for various heavy-quark-mass schemes.  
}
\label{fig:charmpred}
\end{figure}

The combined $F_2^{c\bar{c}}$ data can help to reduce the 
uncertainty on PDFs coming from the choice of heavy-quark-scheme 
and the value of the charm mass input to these schemes.
Fig.~\ref{fig:charmpred} compares the $\chi^2$, 
as a function of the charm mass, for a fit which includes charm data 
(top right) to that for the HERAPDF1.0 fit (top left) when using the 
Thorne-Roberts (RT) variable-flavour-number (VFN) scheme. 
However, this scheme is not unique,
specific choices are made for threshold behaviour. In Fig.~\ref{fig:charmpred} 
(bottom left) the $\chi^2$ profiles for the standard and the 
optimized versions (optimized for smooth 
threshold behaviour) of this scheme are compared. 
The same figure also compares the alternative ACOT VFN 
schemes and the Zero-Mass VFN scheme. Each of these schemes 
favours a different value for the charm quark mass, and the fit to the data 
is good for all the heavy-quark-mass schemes excepting 
the zero-mass scheme. 
Each of these schemes can also be used to predict $W$ 
and $Z$ production for the LHC and their predictions for $W^+$ are shown in 
Fig.~\ref{fig:charmpred} as a function of the charm quark mass (bottom right). 
If a particular
value of the charm mass is chosen then the spread of predictions is as large as
$\sim7\%$.
However this spread is considerably reduced $\sim 1\%$ if each heavy quark 
scheme is used at its own favoured
value of the charm quark-mass.  Furher details of this study are 
given in ref.~\cite{charmcomb}.

\begin{figure}[tbp]
\vspace{-1.0cm} 
\begin{center}
\begin{tabular}{cc}
\includegraphics[width=0.42\textwidth]{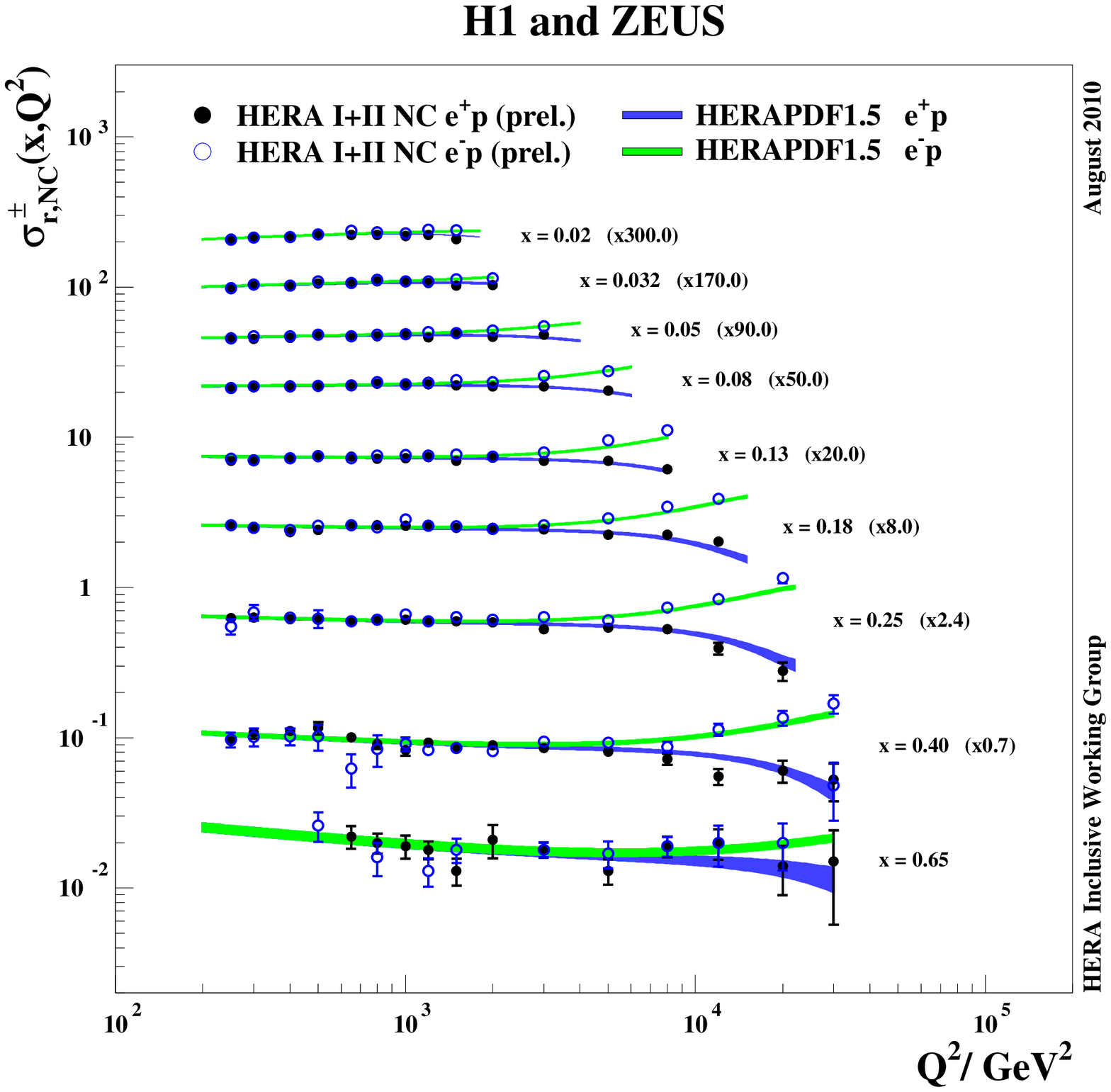} &
\includegraphics[width=0.42\textwidth]{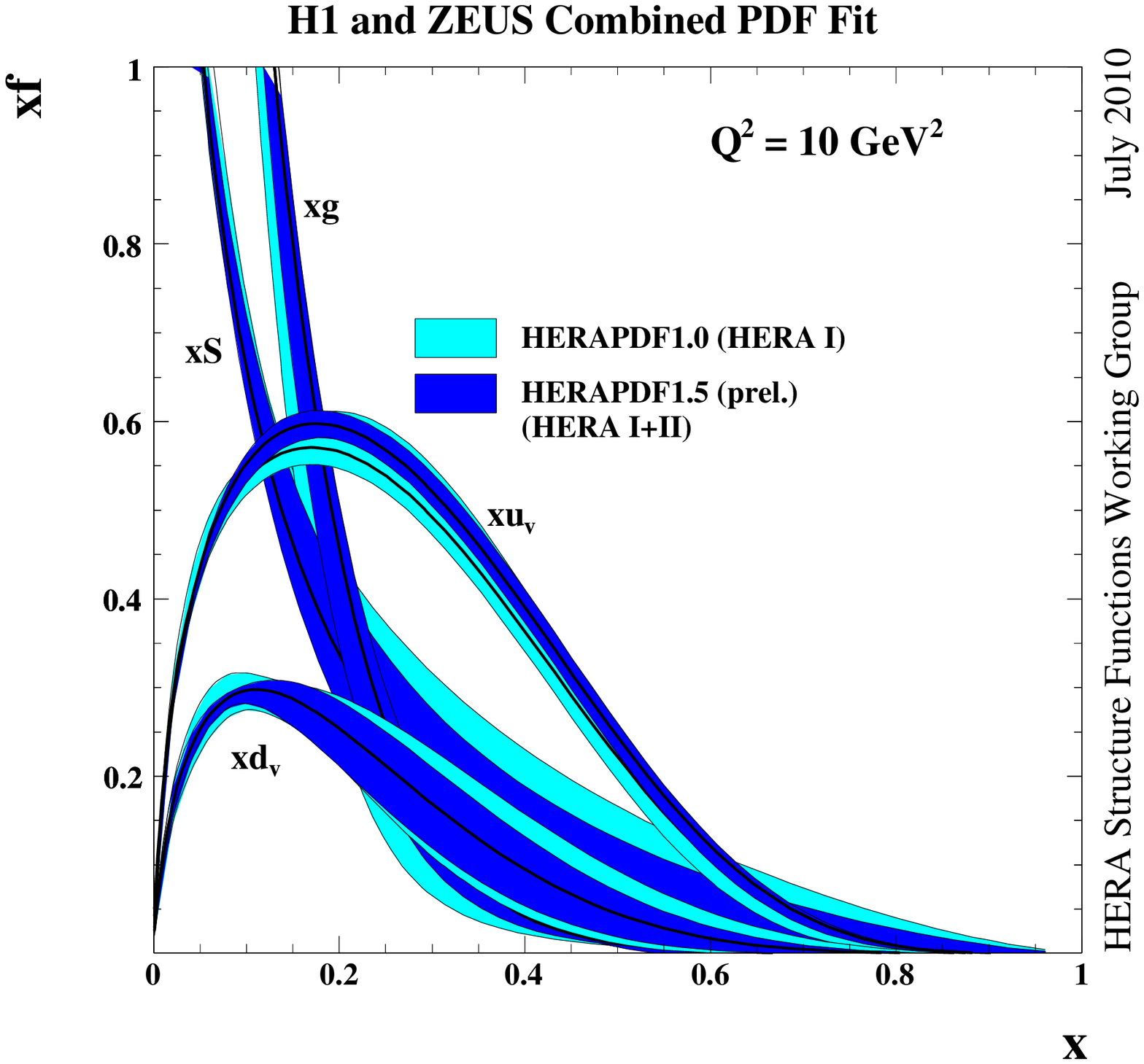}
\end{tabular}
\caption{Left: HERA combined data points for the NC $e^{\pm}p$ cross-sections 
as a function of $Q^2$ in bins of $x$, for data from the HERA-I and II run 
periods. The HERAPDF1.5 fit to these data is also shown on the plot.
Right: Parton distribution functions from HERAPDF1.0 and HERAPDF1.5; $xu_v$, 
$xd_v$,$xS=2x(\bar{U}+\bar{D})$ and $xg$ at $Q^2=10~$GeV$^2$.}
\label{fig:herapdf15}
\end{center}
\end{figure}

The HERA-II data have been combined with the HERA-I data to yield an inclusive 
data set wih improved accuracy at high $Q^2$ and high $x$~\cite{highq2}. 
 This new data set is used as the 
sole input to  a PDF fit called HERAPDF1.5~\cite{herapdf15} 
which uses the same formalism 
and assumptions as the HERAPDF1.0 fit. 
Fig.~\ref{fig:herapdf15} (left) shows the combined data for $NC$ $e^{\pm}p$ 
cross-sections with the HERAPDF1.5 fit superimposed. The parton distribution
functions from HERAPDF1.0 and HERAPDF1.5 are compared in 
Fig.~\ref{fig:herapdf15} (right). The improvement in precision at high $x$ 
is clearly visible.

\begin{figure}[htb]
\begin{tabular}{cc}
\includegraphics[width=0.45\textwidth]{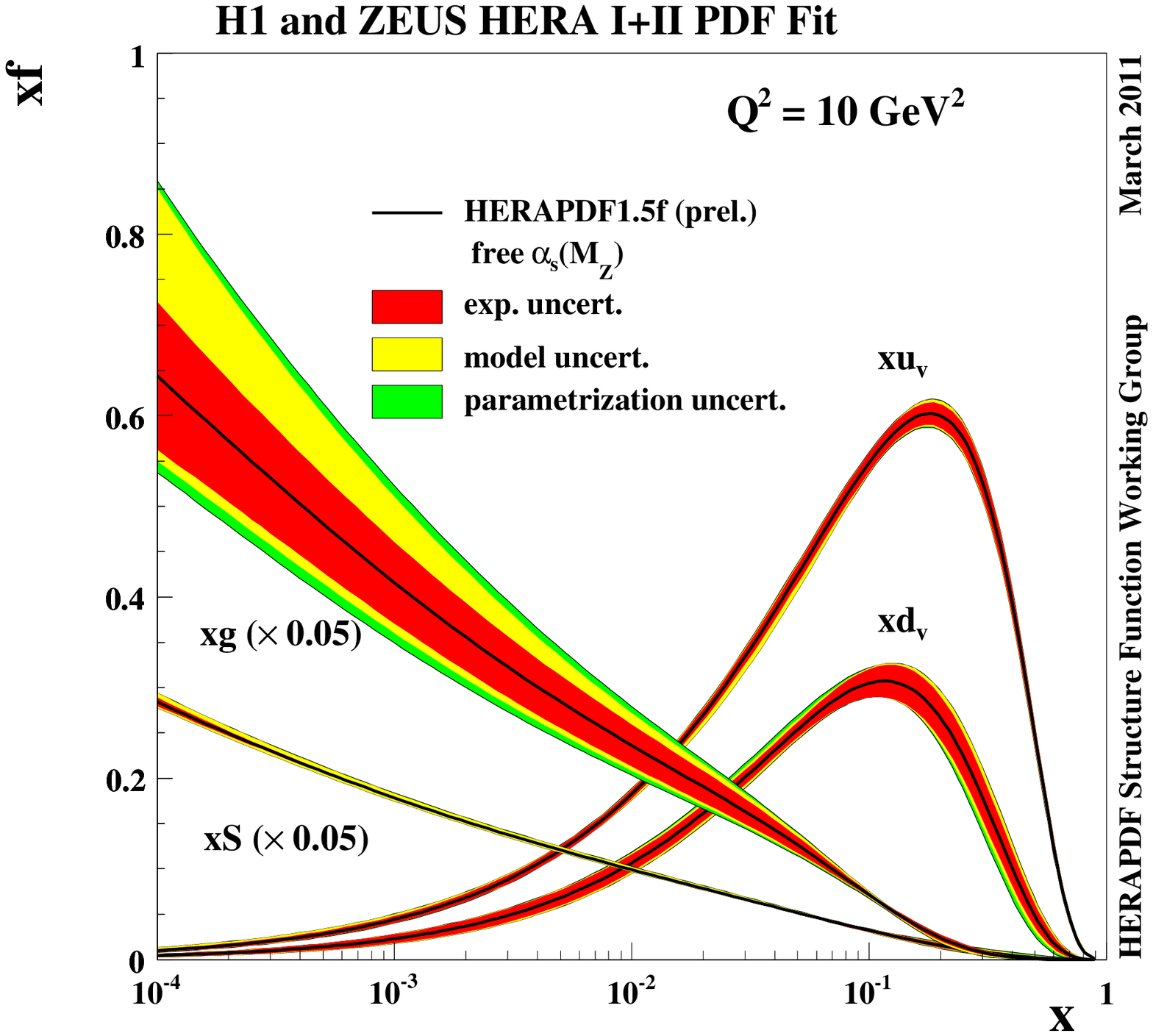} &
\includegraphics[width=0.45\textwidth]{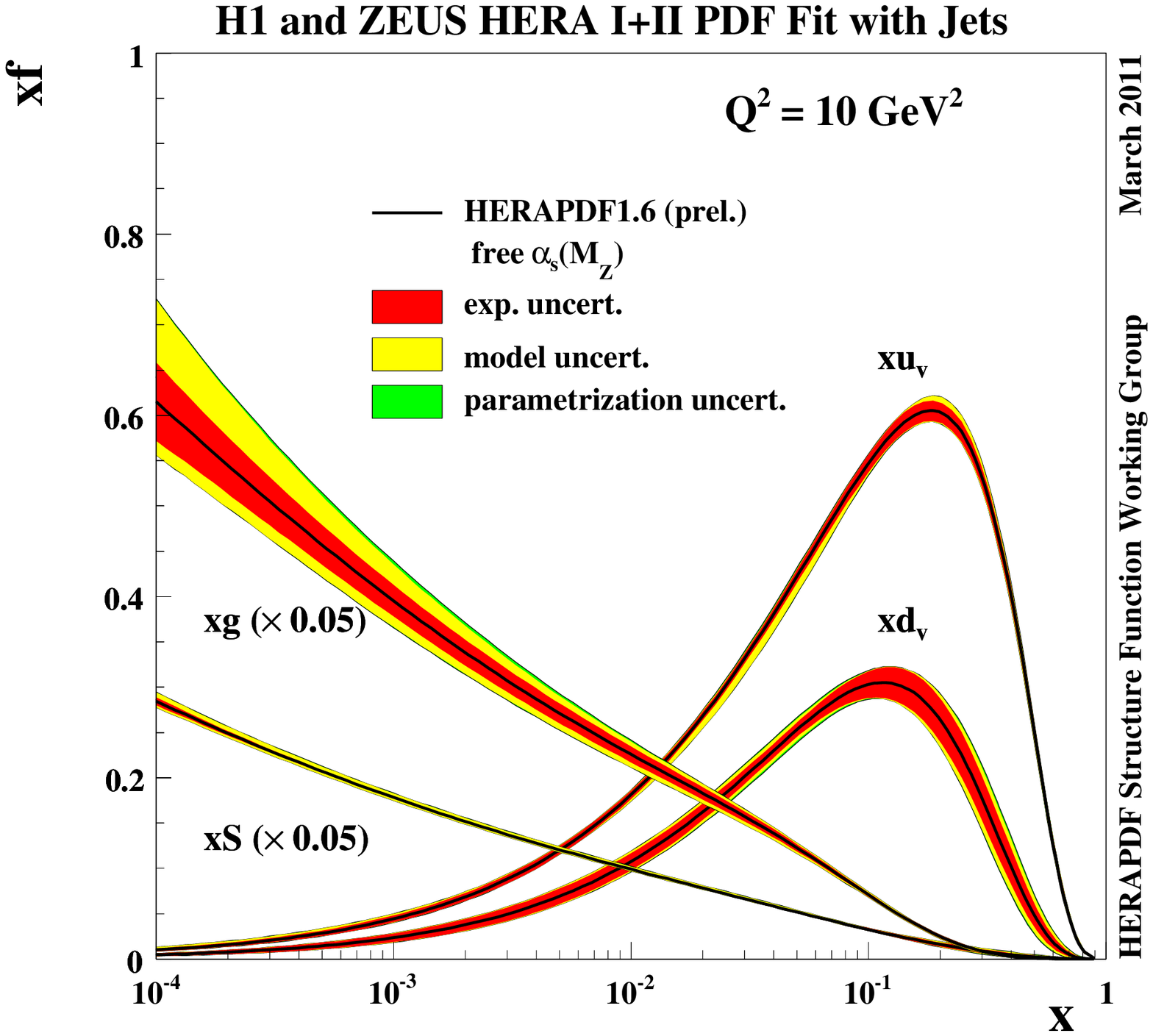}
\end{tabular}
\caption {The parton distribution functions
 $xu_v,xd_v,xS=2x(\bar{U}+\bar{D}),xg$, at $Q^2 = 10$~GeV$^2$, from 
HERAPDF1.5f and HERAPDf1.6, both with $\alpha_S(M_Z)$ 
treated as a free parameter of the fit.
The experimental, model and parametrisation 
uncertainties are shown separately. The gluon and sea 
distributions are scaled down by a factor $20$.
}
\label{fig:jetnojetalph}
\end{figure}

The HERAPDF1.5 analysis has been extended to include HERA inclusive jet data. 
The new PDF set which 
results is called HERAPDF1.6~\cite{herapdf16}.
The gluon PDF contributes only indirectly to the 
inclusive DIS cross sections. However, 
measurements of jet cross sections can provide a direct determination of the 
gluon density.
For these fits the HERAPDF1.5 parametrisation of the gluon distribution $xg$ is 
extended to include a term such that the NLO gluon may become 
negative at low $x$ and low $Q^2$ (however it does not do so in the kinematic 
region of the HERA data). Furthermore,
the low-$x$ slopes of $u$ and $d$ valence PDFs are no longer 
required to be equal.
This more flexible parametrisation is called HERAPDF1.5f. The extra 
flexibility does not change the NLO PDFs central values significantly, and 
the uncertainties are not much increased- the largest increase is in the 
low-$x$ gluon uncertainty, which 
increases from $\sim 8\%$ to $\sim 10\%$ at $Q^2 = 10~$GeV$^2$, $x=10^{-4}$ -  
since HERAPDF always evaluated parametrisation 
uncertainties and model uncertainties as well as purely experimental 
uncertaintes on the PDFs. When the jet data are added the HERAPDF1.6 PDFs are 
also similar, there is no tension between the jet data and the inclusive data. 
The impact of the jet data is seen when 
$\alpha_S(M_Z)$ is allowed to be a free parameter of the fit. 
Fig.~\ref{fig:jetnojetalph} shows the PDFs
for HERAPDF1.5f and HERAPDF1.6, each with $\alpha_S(M_Z)$ left 
free in the fit. It can be seen that without jet data the uncertainty on 
the gluon PDF at low $x$ is large. This is because there is a strong 
correlation between the low-$x$ shape of
the gluon PDF and $\alpha_S(M_Z)$.  However once jet data are 
included the extra information on gluon induced processes reduces this 
correlation and the resulting 
uncertainty on the gluon PDF is not much larger than it 
is for fits with $\alpha_S(M_Z)$ fixed.
The value of $\alpha_s(M_Z)$ extracted from the HERAPDF1.6 fit is:

$
\alpha_S(M_Z) = 0.1202 \pm 0.0013(exp) \pm 0.0007(model/param) \pm 0.0012 (had) +0.0045/-0.0036(scale)
$
Model and parametrisation uncertainties on $\alpha_S(M_Z)$ are estimated
in the same way 
as for the PDFs and the uncertainties on the hadronisation 
corrections applied to the jets are also evaluated. 
The scale uncertainties are estimated by 
varying the renormalisation and factorisation scales chosen in the jet 
publications by a factor of two up and down. The dominant contribution to the 
uncertainty comes 
from the jet renormalisation scale variation. 

\begin{figure}[tbp]
\vspace{-1.0cm}
\begin{tabular}{cc}
\includegraphics[width=0.45\textwidth]{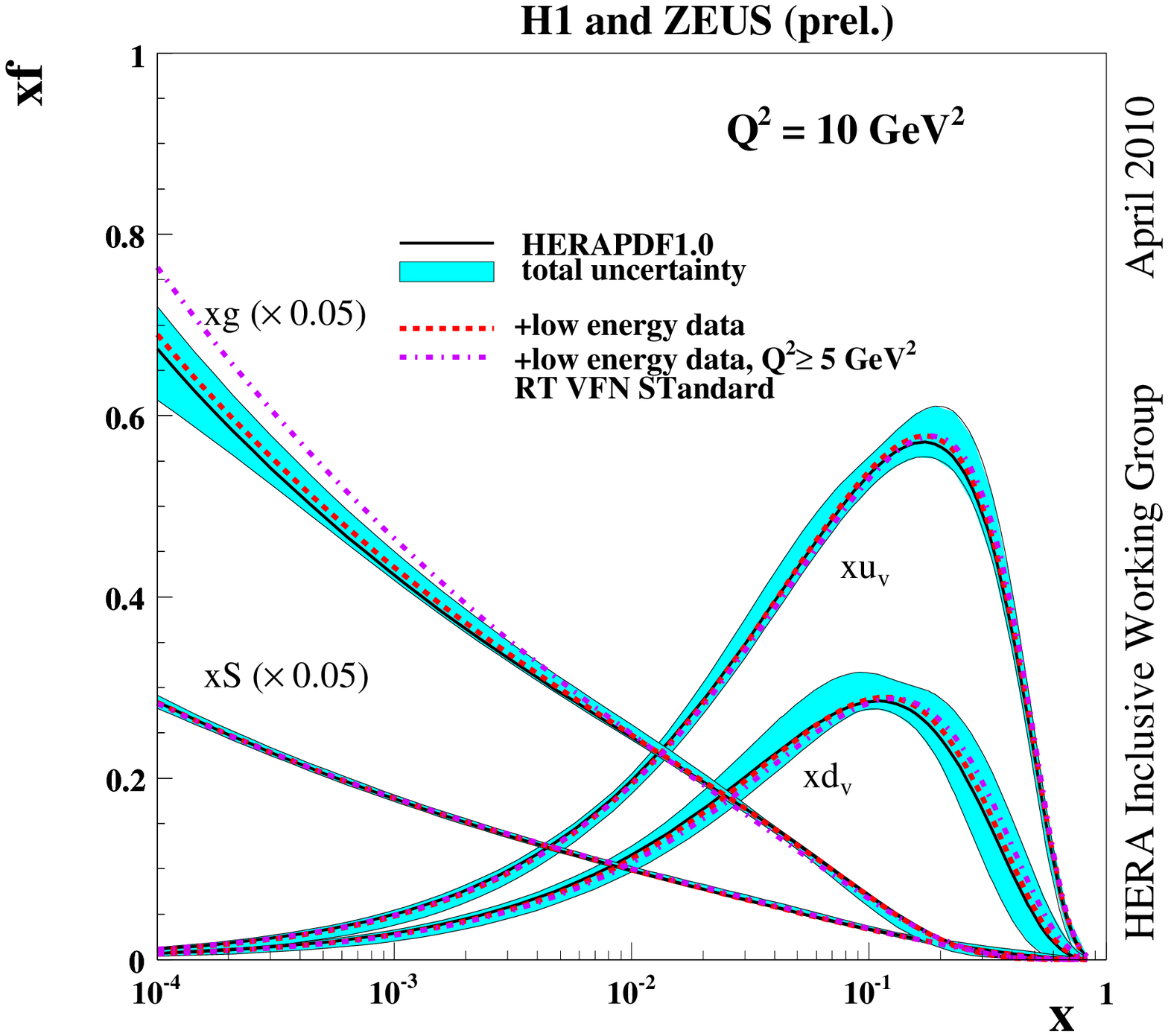} &
\includegraphics[width=0.45\textwidth]{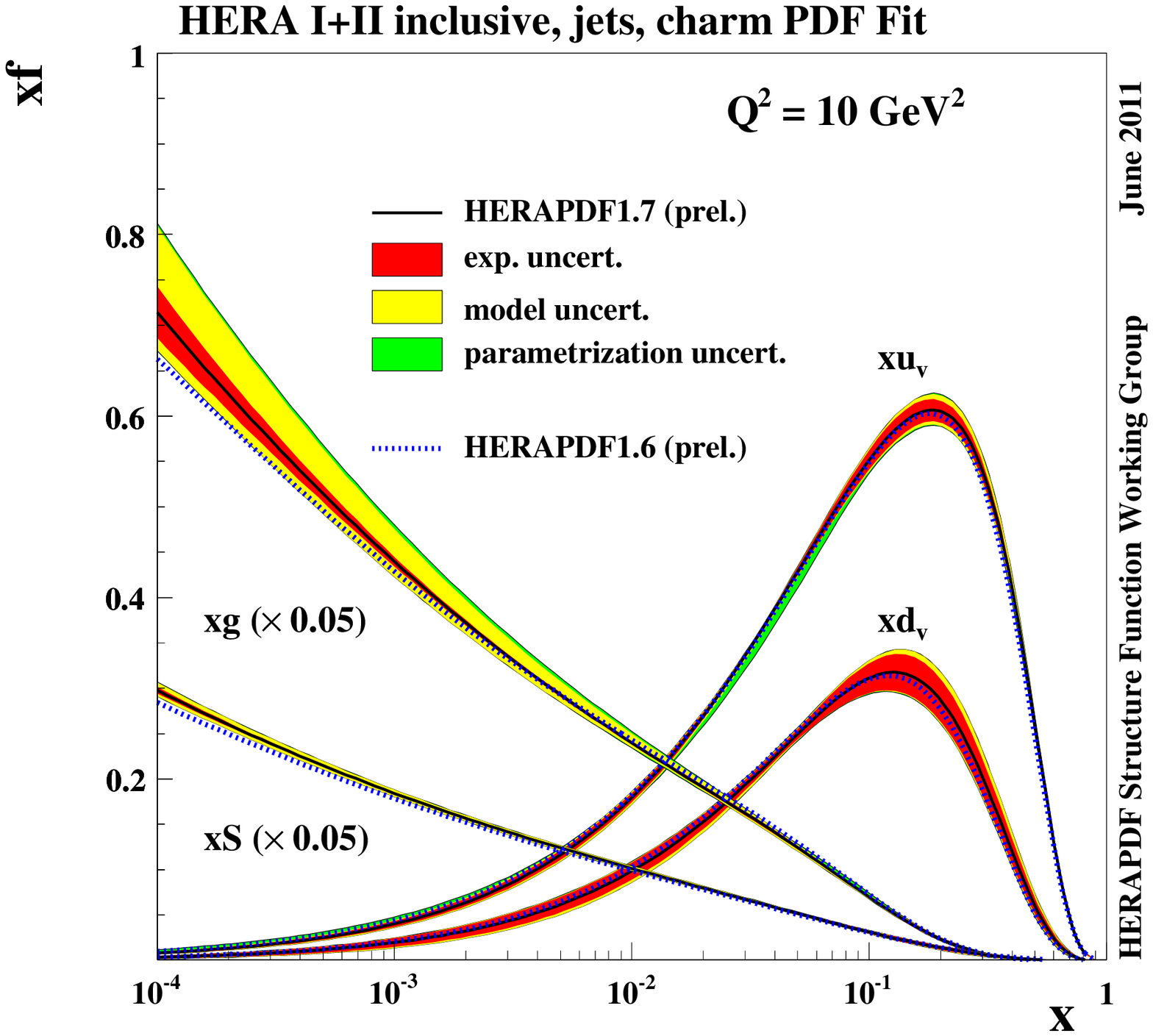}
\end{tabular}
\caption {Left: PDFs from fit to HERA-1 data plus combined low energy data, 
for two values of the minimum $Q^2$ cut, compared to HERAPDF1.0. Right: PDFs 
from HERAPDF1.7, which includes charm data, jet data and low energy data as well 
as the HERA-I and II high energy inclusive data
}
\label{fig:lowenergy17}
\end{figure}

In 2007 NC$e^+p$ data were taken at two lower values of the proton beam energy 
in order to determine the longitudinal structure function $F_L$. Some of the 
data sets from H1 and ZEUS have been combined and used as input to the PDF 
fits together with the HERA-I data. The resulting PDFs are compared with 
HERAPDF1.0 in Fig.~\ref{fig:lowenergy17}. The low energy data are sensitive 
to the
choice of the minimum $Q^2$ of data entering the fit (standard cut 
$Q^2 > 3.5~$GeV$^2$). If this cut is raised to $Q^2 > 5~$GeV$^2$ a 
steeper gluon results. This sensitivity is also seen if a $x$ cut 
($x > 0.0005$) or a saturation inspired cut $Q^2 > 0.5 x^{-0.3}$ is made.
However, these variations are comparable to the uncertainty at low-$x$ 
which is covered by the HERAPDF1.5f flexible parametrisation. 
This parametrisation has been used for a fit which includes all inclusive 
HERA data from HERA-I and II, charm data, low energy data and jet data. For this 
fit the central settings of the HERAPDF1.5f(6) fits are modified such that 
$\alpha_S(M_Z)=0.119$ and $m_c=1.5$ is used 
together with the RT-optimized heavy quark scheme 
following the preferences of the jet and charm data.
This fit, called HERAPDF1.7, is illustrated on the right hand side of 
Fig.~\ref{fig:lowenergy17}

\begin{figure}[tbp]
\begin{tabular}{cc}
\includegraphics[width=0.45\textwidth]{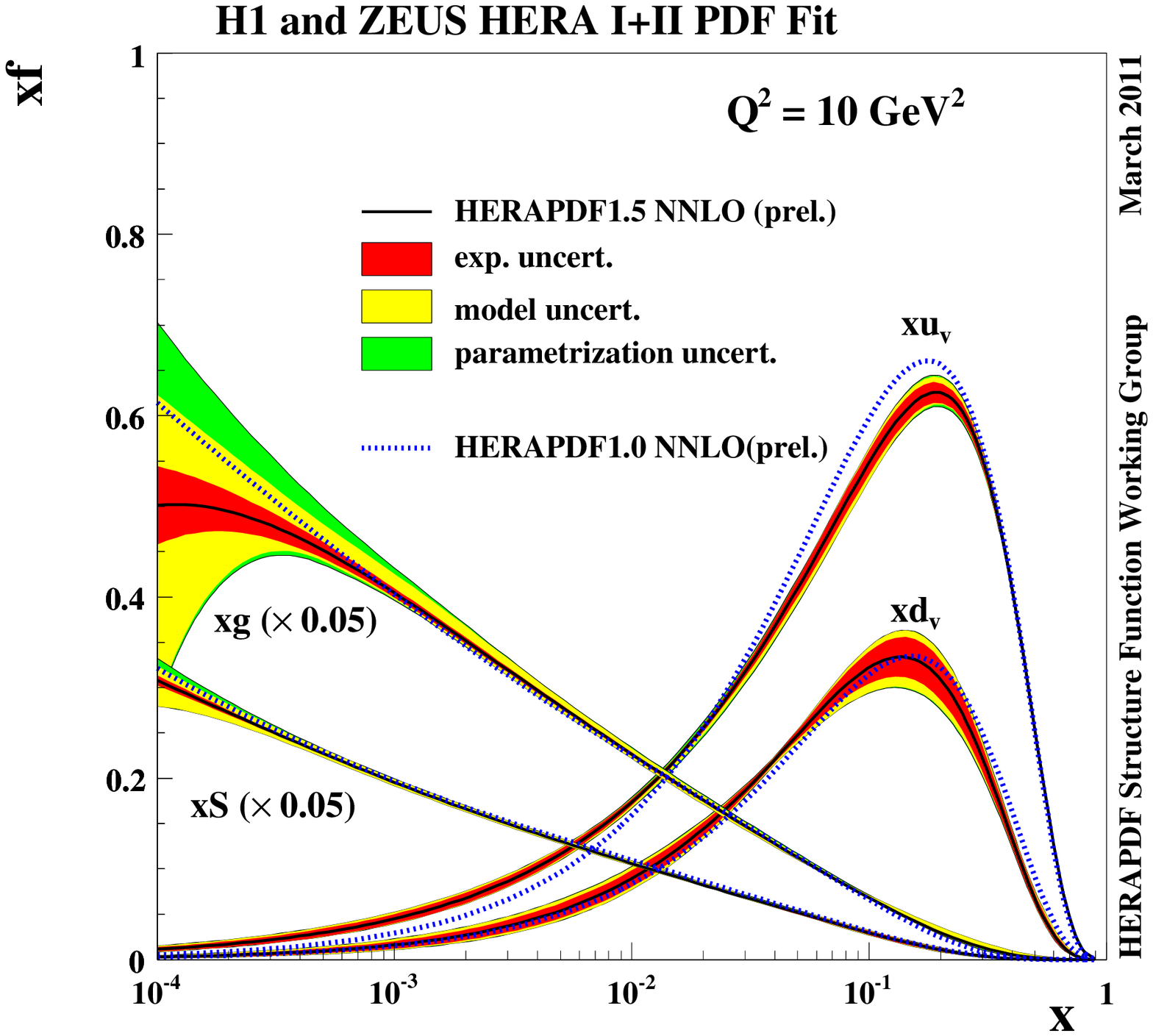} &
\includegraphics[width=0.45\textwidth]{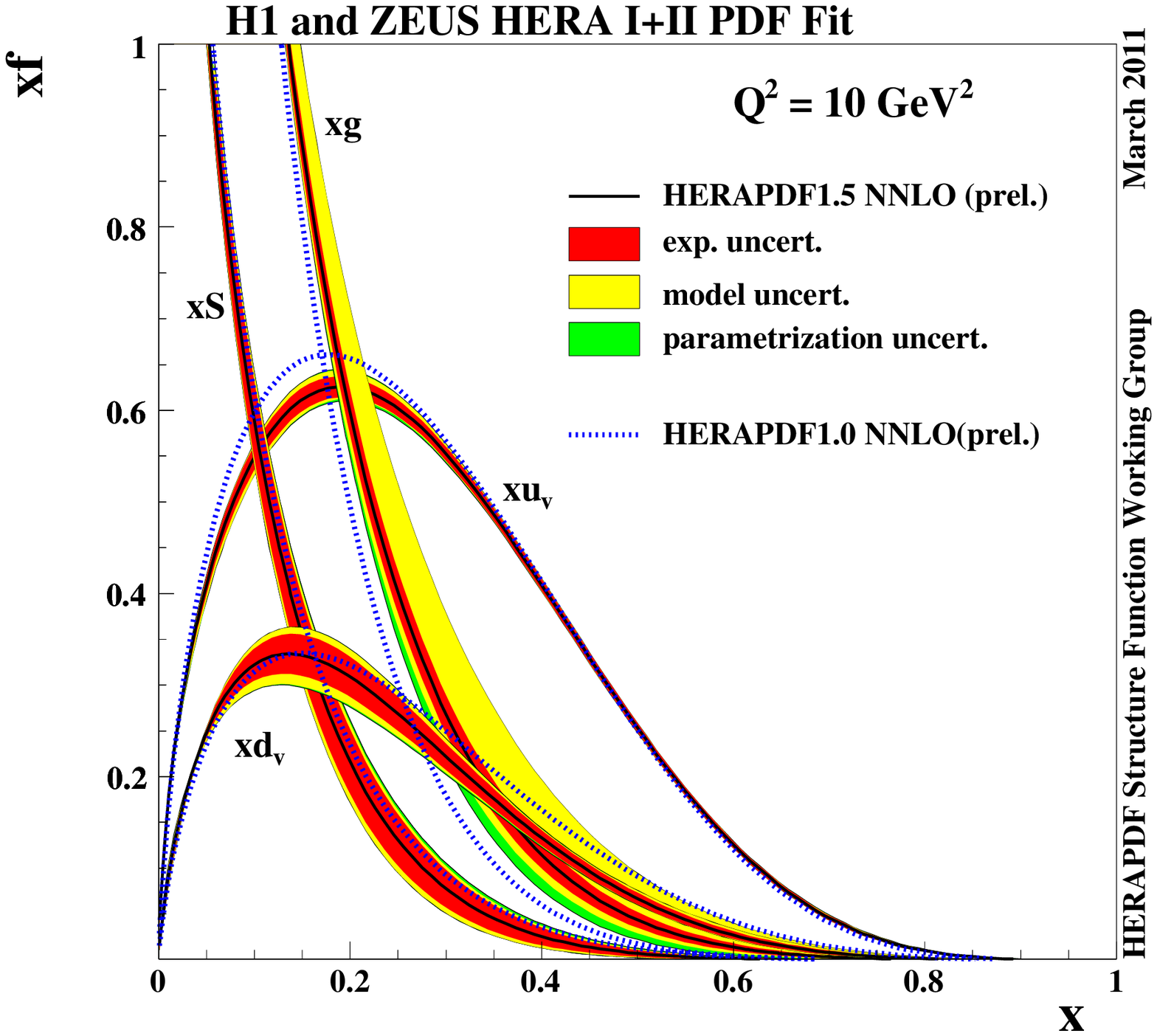}
\end{tabular}
\caption {HERAPDF1.5NNLO PDFs compared to HERAPDF1.0NNLO PDFS on log 
and linear $x$ scales. 
}
\label{fig:nnloherapdf}
\end{figure}

A preliminary NNLO extraction HERAPDF1.0 NNLO was presented in 2010 but this 
has been updated to
HERAPDF1.5NNLO~\cite{herapdf15nnlo}, with full accounting for experimental, 
model and parametrisation uncertainties, and using the extended form of the 
parametrisation.
Fig.~\ref{fig:nnloherapdf} compares the 
HERAPDF1.5NNLO with HERAPDF1.0 NNLO for $\alpha_s(M_Z)=0.1176$, 
which is our recommended value for $\alpha_S(M_Z)$ at NNLO. 
The HERAPDF1.5 NNLO fit 
has a signficantly harder high-x gluon.  

\begin{figure}[tbp]
\begin{tabular}{cc}
\includegraphics[width=0.4\textwidth]{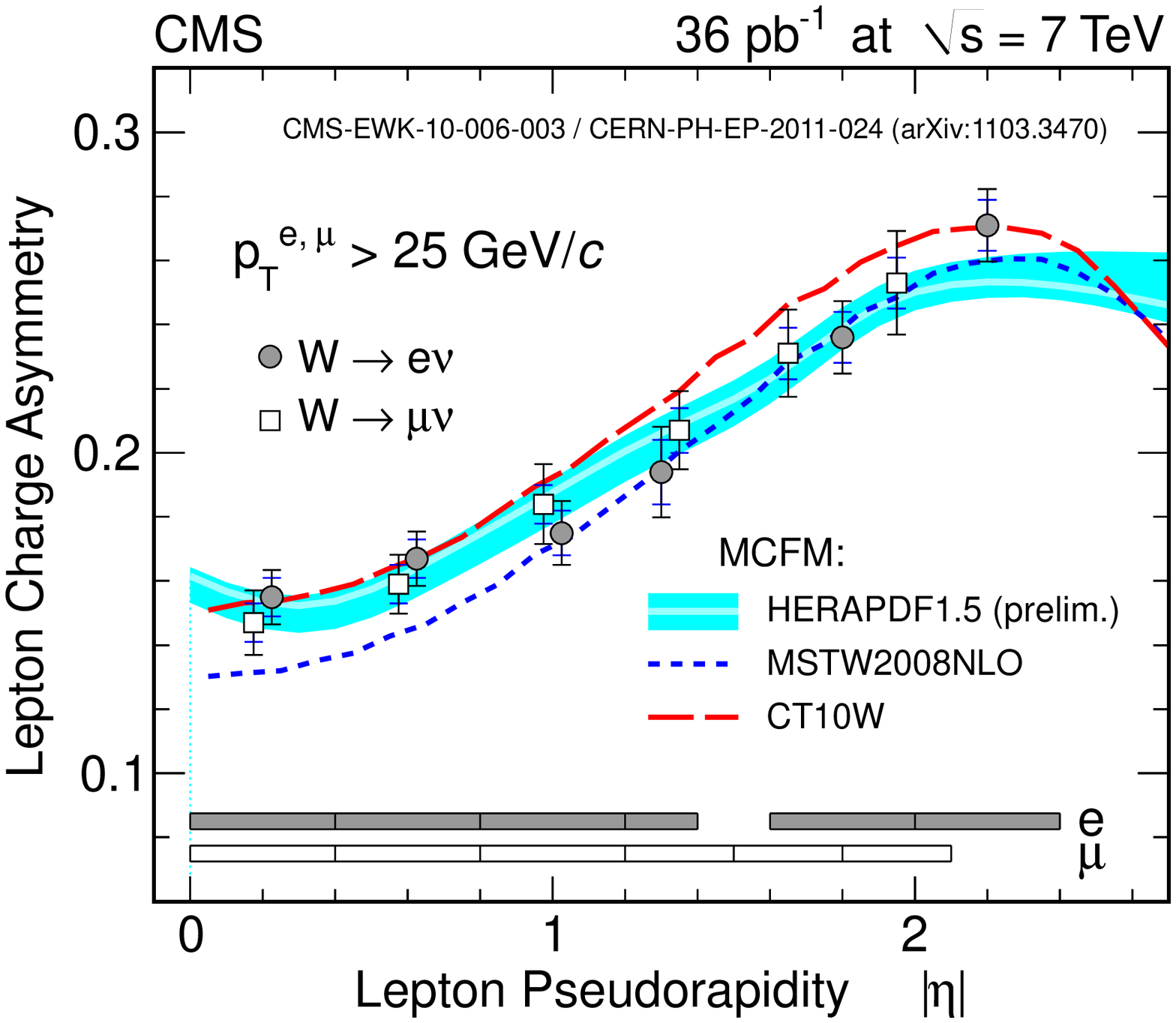} &
\includegraphics[width=0.4\textwidth]{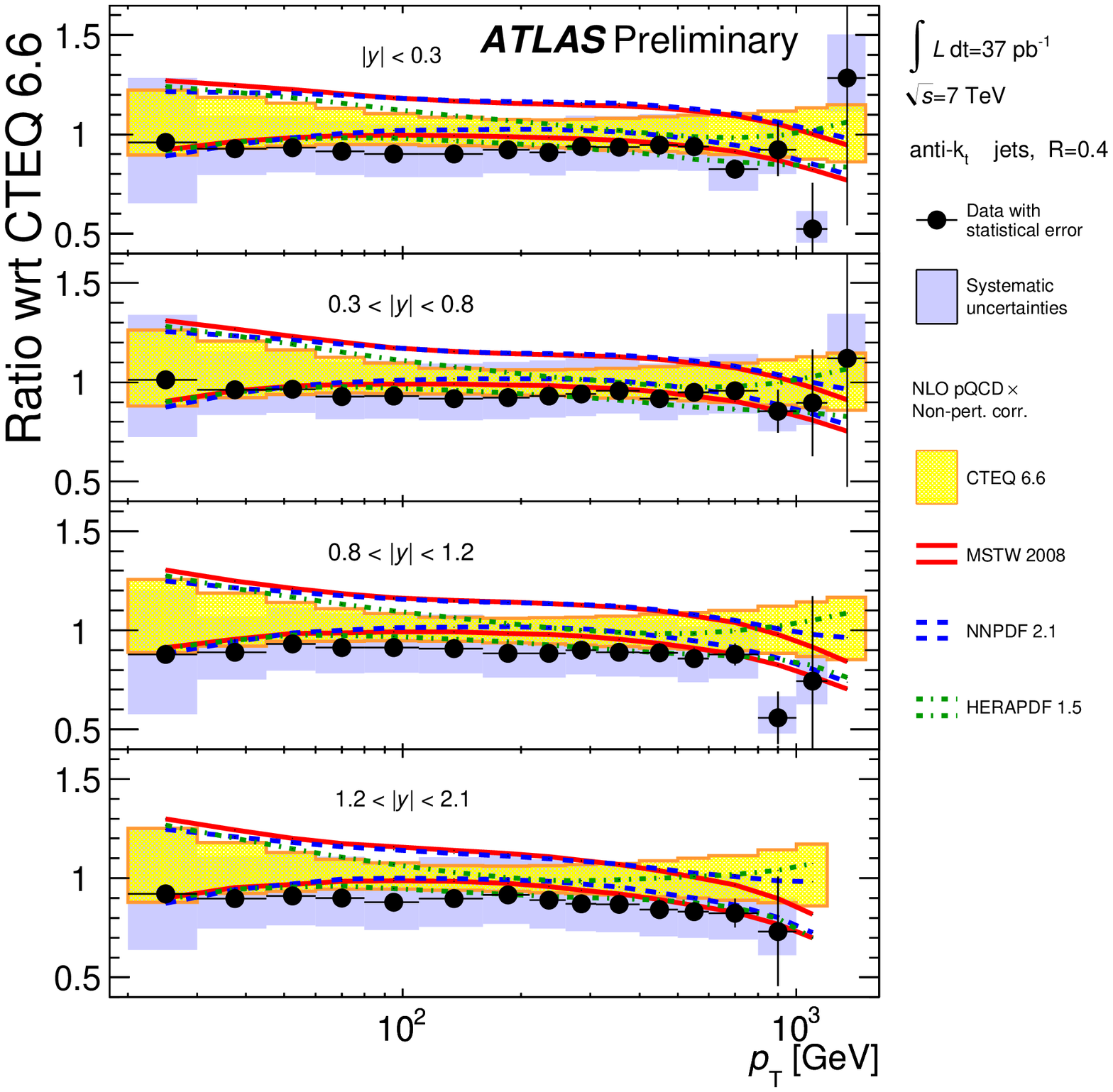}
\end{tabular}
\caption {Left: HERAPDF1.5 predictions for LHC $W$-lepton asymmetry data from CMS. Right ATLAS jet data in the central region in ratio to the predictions of CTEQ6.6 and compared to other PDF predictions.
}
\label{fig:LHC}
\end{figure}

Finally the HERAPDFs have been successfully 
confronted with both Tevatron and LHC data on 
$W,Z$ and jet production. 
Fig.~\ref{fig:LHC} show comparisons of the HERAPDF1.5 NLO
 predictions to the early 
LHC data on the W asymmetry from CMS and comparisons of various 
PDFs to the ATLAS inclusive jet data. 

\section{Summary}
The addition of new data sets to HERAPDF1.0 has given new information. The 
addition of inclusive data from HERA-II running, as used for the HERAPDF1.5,
 has improved the precision of 
PDFs at high $x$ and $Q^2$. The addition of charm data helps to fix the charm 
mass and the heavy quark scheme. The addition of jet data, as used for HERAPDF1.6, yields a competitive 
measurement of $\alpha_s(M_Z)$. 
The PDF analysis has been extended from NLO to NNLO. The HERAPDF has been 
successfully confronted with Tevatron and LHC data on $W,Z$ production and 
jet production. The HERAPDF1.5 NLO and NNLO PDFs are available on LHAPDF
(LHAPDFv6.8.6, http:projects.hepforge.org/lhapdf)

\end{document}